\documentstyle[11pt,newpasp,twoside,epsf]{article}
\def\etal{et al.}
\def\hii{H{\sc ii}}

\def\ii{\'{\i}}
\def\micron{$\mu$m}

\def\msun{M$_{\odot}$}

\def\halpha{\ifmmode {\rm H{\alpha}} \else $\rm H{\alpha}$\fi}
\def\hbeta{\ifmmode {\rm H{\beta}} \else $\rm H{\beta}$\fi}
\def\heic{He\,{\sc i} $\lambda$5876}

\def\heii{\ifmmode {{\rm He{\sc ii}} \lambda 4686} \else {He\,{\sc ii} $\lambda$4686}\fi}
\def\Heii{He\,{\sc ii}}

\def\oia{[O\,{\sc i}] $\lambda$6300}

\def\oiii{[O\,{\sc iii}]}

\def\oiiic{[O\,{\sc iii}] $\lambda$4363}

\def\civ{C\,{\sc iv} $\lambda$5808}
\begin{document}
\title{Massive Star Forming Regions: from diagnostic tools to derived properties}

\author{Daniel Schaerer} 
\affil{Laboratoire d'Astrophysique, Observatoire Midi-Pyr\'en\'ees, 
       14, Av. E. Belin, F-31400, Toulouse, France
      (schaerer@ast.obs-mip.fr)}

\begin{abstract}
The current state-of-the-art of multi-wavelength diagnostic tools 
(evolutionary synthesis, photoionisation models) for massive star 
forming regions (\hii\ regions, starbursts, etc.)
and some of their input physics (especially model atmospheres) is 
reviewed.
Analysis of stellar populations based on integrated spectra from both
stellar features and nebular emission lines from the UV to IR are summarised.
We stress the importance of ``template'' studies at various scales
(from individual stars to well studied galaxies) and various 
wavelengths, to understand the processes operating in massive 
star forming regions and to reliably derive their properties.

\end{abstract}

\section{Introduction}
In a variety of objects ranging from stellar clusters, \hii\ regions,
over dwarf and irregular galaxies, to full-fledged starburst (SB) galaxies and
distant Lyman-break galaxies, massive star formation plays a major role
in the budget of radiative, mechanical and chemical output. 
Indeed, when present, such stars (mostly OB and Wolf-Rayet stars) dominate
the light at UV--optical wavelength, they are responsible for the ionisation
of the ISM, and they also dominate (together with SN) mechanical feedback 
processes.

The present review deals with diagnostic techniques and the properties of
objects of the types mentioned above, for which the ``generic'' term
massive star forming regions (MSFR) will be used subsequently.
As such we are mostly limited to objects where the properties of
integrated populations (as opposed to resolved stellar populations;
cf.\ Gallart, Grebel, these proceedings) are observed. 
The aim is to review the current state-of-the-art of various tools
(evolutionary synthesis, photoionisation models) and some of their
inputs (especially ionising fluxes), to assess their main successes
and failures (and the lessons to be learned from that), and to 
summarise some of the main properties of stellar populations in various
MSFRs.


\section{Radiative output from massive stars}
Schematically the main observed UV to far-IR features of integrated 
stellar populations can be grouped as follows:
\begin{enumerate}
\item {\em Continuum:} From the UV to $\sim$ 4 \micron\ (approx.\ L
  or M bands) the continuum
  is usually dominated by stellar light, except in very young objects
  (e.g.\ strong emission line (EL) objects) where continuous nebular 
  emission may be 
  non-negligible and may in particular affect observed colours.
  At longer wavelength dust emission usually starts to dominate
  the continuum emission.
\item {\em Stellar signatures:}
  Due to the increasing nebular and dust emission towards the near-IR,
  stellar signatures are preferentially detected at UV -- optical --
  K band wavelengths.
  Among the strongest stellar features are: UV P-Cygni lines
  (C~{\sc iv} $\lambda$1550, Si~{\sc iv} $\lambda$1400, $\ldots$)
  emitted in the stellar winds of OB stars, broad emission lines
  (\heii, \civ) from Wolf-Rayet (WR) stars, H and He absorption lines
  from OBA stars in the optical, and various absorption features
  (e.g.\ Ca~{\sc ii} triplet, TiO features in the optical, CO absorption
  bands in K band) from cool giants and supergiants.
\item {\em Nebular lines:} Relatively few and faint lines are in the UV,
  opposed to numerous H and He recombination lines and forbidden lines 
  from abundant metals (N, O, S etc.) in the optical. 
  A rich spectrum of H recombination lines and fine structure 
  lines is observed from the near to far IR 
  (Ne, S, Ar: $\sim$ 3-40 \micron; C, N, O $\ga$ 
  40 \micron). The emission originates mostly from ionised gas 
  (``\hii''\ regions in the former case; the latter also includes lines 
  emitted in photo-dissociation regions (PDR; e.g.\ [C~{\sc ii}] 158 \micron).
\end{enumerate}

Subsequently we shall summarise the current state-of-the-art of 
analysis of stellar populations based on stellar signatures
(Sect.\ 3) and through nebular lines (Sect.\ 4).

\section{Starbursts in the UV and optical: stellar features}
In recent years analysis of UV spectra have mostly been performed
by the groups of Leitherer, Heckman, Robert, Meurer, Conti, 
Gonz\'alez-Delgado and other co-workers, Mas-Hesse \& Kunth and
others.
Most of the observations are based on HST (GHRS or STIS), HUT and 
IUE (FUSE data upcoming) of nearby starbursts. 
Restframe UV spectra of Lyman-break galaxies (e.g.\ Lowenthal \etal\ 1997,
Pettini \etal\ 2000) now allow to extend such studies to high-redshift 
star forming galaxies.
The {\em Starburst99} evolutionary synthesis models (Leitherer \etal\
1999) used for the interpretation of the data include for this wavelength 
range empirical IUE high-resolution spectral libraries of Galactic O and 
WR stars (+ B stars: de Mello \etal\ 2000); low-resolution spectra are used
in the Mas-Hesse \& Kunth (1999) models.
Work by several groups is in progress to include libraries of low metallicity 
stars. 

Most studies where this UV technique has been applied yield similar results
which can, ``on average'', be summarised as follows (e.g.\ Gonz\'alez Delgado 
\etal\ 1998, Leitherer 1999b):
{\em 1)} the presence of the UV wind features indicates young bursts 
   (ages $\la$ 10--20 Myr);
{\em 2)} often it is difficult to distinguish between constant SF or 
  instantaneous bursts;
{\em 3)} the observations are generally compatible with a ``normal'' Salpeter
  IMF extending to high masses ($M_{\rm up} \sim$ 60--100 \msun).
Extended SF and older ages are e.g.\ favoured for the high redshift
galaxy cB58 (de Mello \etal\ 2000, Pettini \etal\ 2000). Such a difference is
probably simply due a larger mixture of various regions included in the 
aperture for such distant an object.
Some interesting extensions of the UV analysis are the potential use of 
wind lines to estimate the metallicity of these objects (Leitherer 1999a), 
and the simultaneous interpretation of interstellar lines indicative
of the state of the ISM (e.g.\ Gonz\'alez Delgado \etal\ 1998, Heap 2000).

In the optical, studies of the so-called Wolf-Rayet (WR) galaxies 
(see compilation by Schaerer \etal\ 1999b) have provided useful 
insight on massive star populations in starbursts,
in particular because these objects cover a large metallicity range
and, given the nature of WR stars, they represent the best probe of the
upper end of the IMF. Recent reviews on WR galaxies are found in
the IAU Symp.\ 193 (van der Hucht \etal\ 1999), Mas-Hesse (1999), and
Schaerer (1999ab). Here we shall briefly summarise the main results
from the large sample of Schaerer \etal\ (1999a) and Guseva \etal\ (2000).

In general a good agreement is found between the observations and the
synthesis models of Schaerer \& Vacca (1998), with a possible exception
for some very low metallicity objects (Guseva \etal\ 2000, but also 
de Mello \etal\ 1998). This comparison indicates fairly short
time scales of SF (bursts with $\Delta t \la$ 2--4 Myr) for the bulk
of the objects at subsolar metallicity (mostly BCD galaxies). Again
the IMF is compatible with a Salpeter slope and the existence of
high mass stars is required. Similar conclusions are obtained by
Mas-Hesse \& Kunth (1999) and in earlier studies. The detection
of WR stars of both WN and WC subtypes and the deduced WC/WN ratio
provides strong constraints for stellar evolution models 
(see Schaerer \etal\ 1999a).
This work was recently extended to a detailed analysis of five 
metal-rich starbursts which indicates a {\em lower limit} of
$M_{\rm up} \ga$ 30--40 \msun\ for the upper mass cut-off
of the IMF in such environments (Schaerer \etal\ 2000).

New models synthesising the higher order Balmer and He~{\sc i} absorption
lines were constructed by Gonz\'alez Delgado \etal\ (1999ab;
see references therein for earlier work).
These lines are detected in a variety of objects (starbursts, Seyfert 
2, post starbursts etc.) and provide useful information on massive
(OB) star populations as well as later types (A). 
The first applications of these models to starbursts and \hii\ regions
are demonstrated in Gonz\'alez Delgado \etal\ (1999b) and Gonz\'alez 
Delgado \& P\'erez (2000). 
With no doubt this tool will be very useful to unravel massive star 
populations from integrated spectra.

\section{Revealing starburst properties from nebular lines}
Several types of diagnostics can be drawn from measurements
of nebular lines:
\begin{itemize}
\item 
 ``Basic diagnostics'', i.e.\ quantities such as the
 total ionising photon flux, electron densities, temperatures,
 abundance ratios (requiring knowledge of $T_e$ and a correction
 for unobserved ionisation stages), which can be derived from 
 first principles.
\item 
  ``Advanced diagnostics'', such as the $R_{23}$ or $S_{23}$ 
  metallicity indicators (e.g.\ Vilchez \& Esteban 1996, D\'{\i}az
  \& P\'erez-Montero 2000) and stellar temperature indicators
  (softness parameter $\eta^\prime$: cf.\ Vilchez \etal\ 1998;
  HeI/\hbeta: Kennicutt \etal\ 2000;
  [Ne~{\sc iii}]/\hbeta: Oey \etal\ 2000), are based on empirical 
  calibrations and/or photoionisation models.
\item 
Other diagnostics require the use of photoionisation models.
\end{itemize}

In all cases the translation to the physical properties of interest
here, i.e.\ the exciting stellar population, depends on the additional
use of  stellar atmosphere and stellar evolution models as well as on 
evolutionary synthesis models. 
Given the strong sensitivity of the derived parameters to these fundamental
ingredients (stellar models) it is important to verify their reliability.
In the following we shall briefly summarise the current status 
of such tests regarding the ionising fluxes from massive stars.
The latest developments in stellar evolution models are reviewed in
Maeder \& Meynet (2000).

\subsection{Ionising fluxes of O and WR stars: theory and observations}
State-of-the-art atmosphere models for O stars include non-LTE effects,
stellar winds and line blanketing ({\em CoStar models:} Schaerer \& de Koter 1997;
Pauldrach \etal\ 1998) which lead to important modifications
of the ionising spectrum when compared e.g.\ to the popular
plane parallel LTE models of Kurucz. 

The implications of the {\em CoStar} fluxes
on \hii\ regions have been discussed by Stasi\'nska \& Schaerer (1997)
and in various papers relying on these atmosphere models.
Agreement is found for the predicted number of Lyman continuum photons, $N_{\rm Lyc}$,
from studies of nebulae with known stellar content (Oey \& Kennicutt 1997).
However, this does not represent very strong constraints since the observations
provide a priori only a lower limit to $N_{\rm Lyc}$, and for stars with
$T_{\rm eff} \ga$ 40 kK the predicted $N_{\rm Lyc}$ is little model dependent.
Support for the predicted SED of the {\em CoStar} models in the Lyman continuum 
comes from comparisons of optical and IR observations to grids of photoionisation models
(Stasi\'nska \& Schaerer 1997), detailed modeling of two LMC nebulae with
known stellar content and a constrained ionisation parameter (Oey \etal\ 2000), 
and the agreement with the He~{\sc i}/H hardness 
(indicated by He~{\sc i} $\lambda$5876/\hbeta) over a large range of $T_{\rm eff}$ 
shown in Figure 1. 
The work of Oey \etal\ (2000) indicates a possibly 
too hard spectrum in the 41-54 eV range.
The first comparisons of the Pauldrach \etal\ (1998) models with
\hii\ region data shows a similar agreement with He~{\sc i} $\lambda$5876/\hbeta\
(Kennicutt \etal\ 2000). Further detailed comparisons along the lines
of the work of Oey \etal\ (2000) are necessary to provide accurate constraints 
on the atmosphere models. 
The reasonable success of the {\em CoStar} models indicates that such
models can quite confidently be used in the interpretation of MSFR.

\begin{figure}[tb]
\plottwo{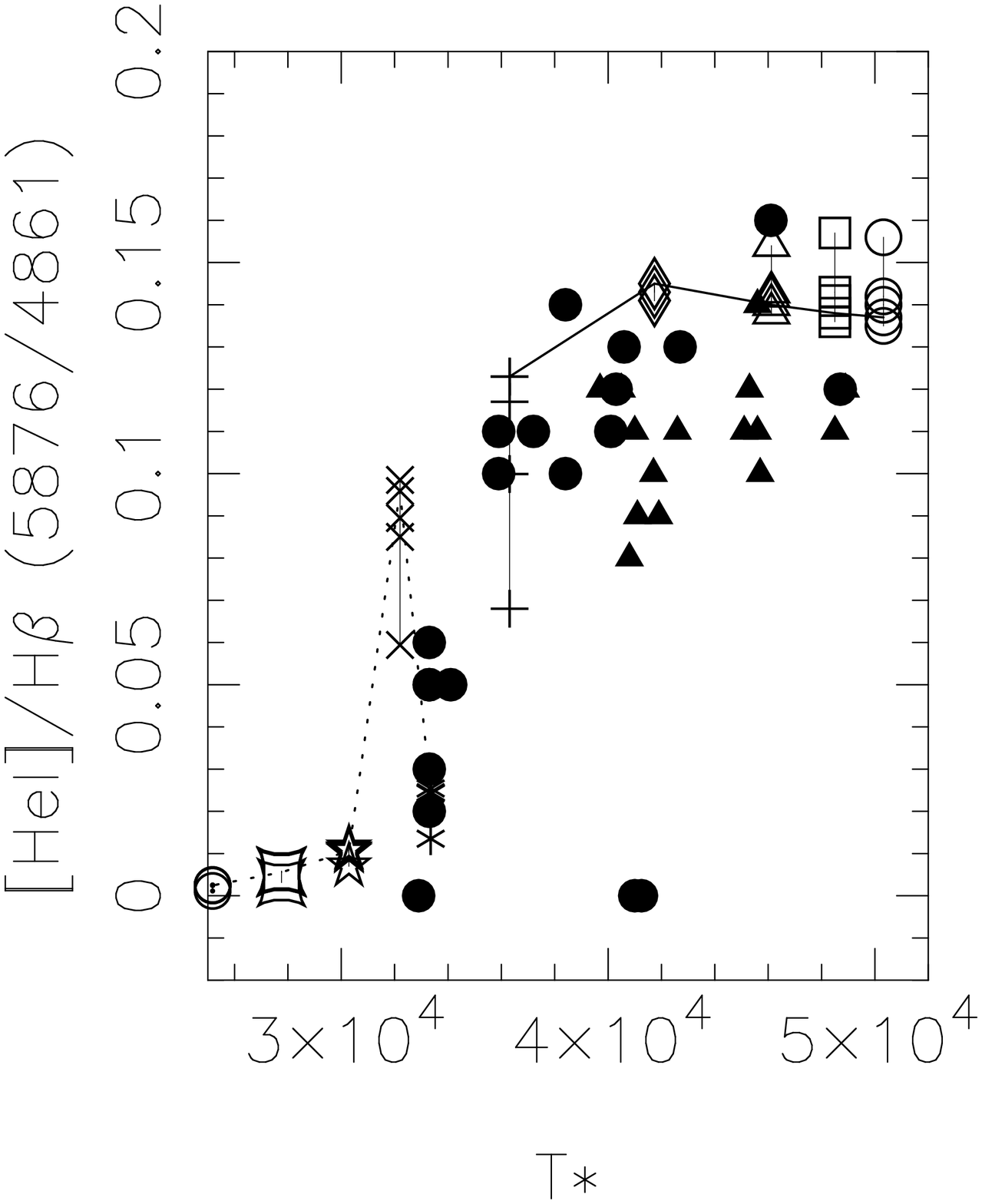}{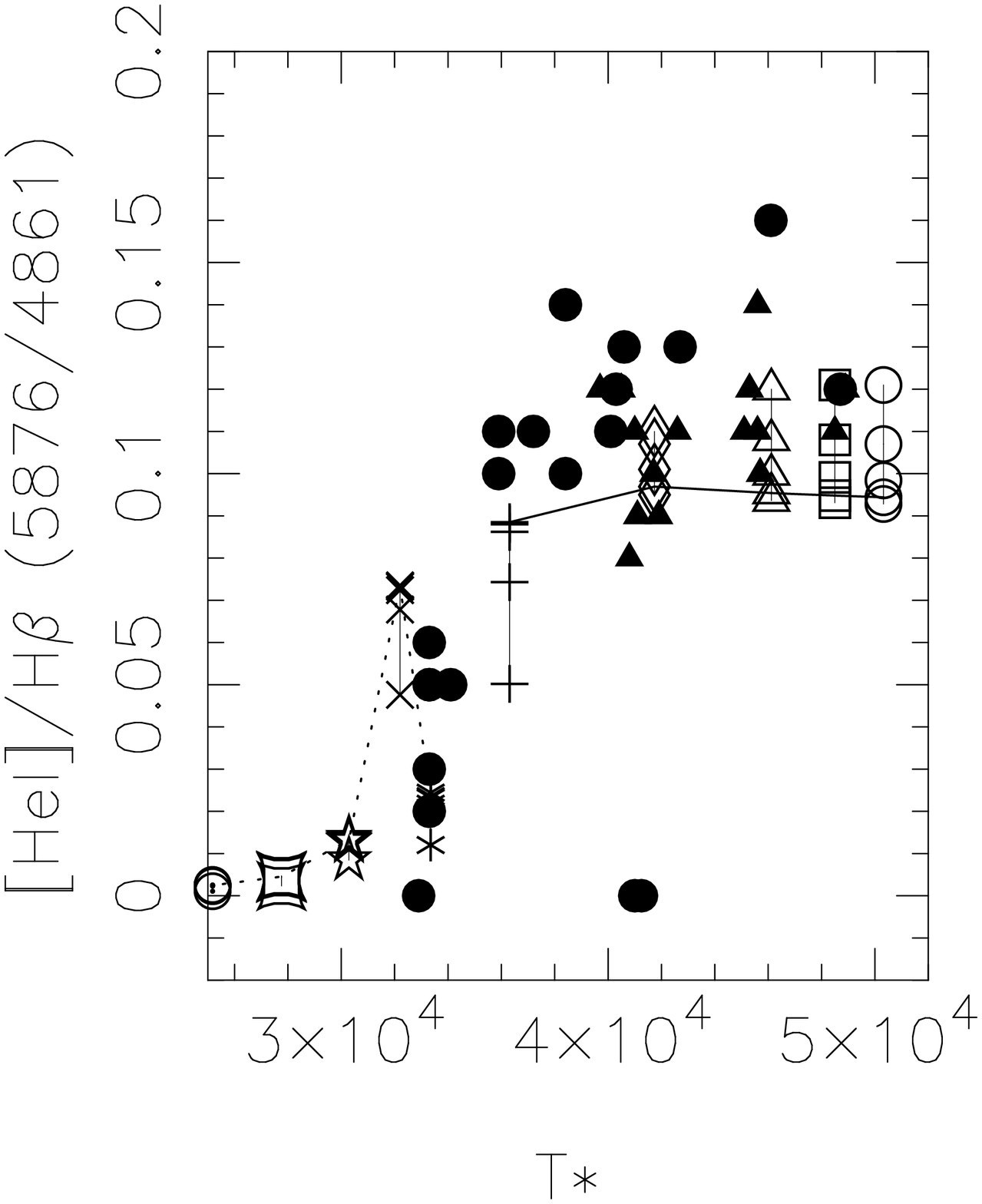}
\caption{Observed and predicted \heic/\hbeta\ ratio versus stellar effective temperature.
Filled symbols: Observations of Galactic and LMC \hii\ regions from Kennicutt \etal\ 
(2000, cf.\ their Fig. 9) plotted versus temperature from using the Vacca \etal\ (1996) 
temperature scale.
Open symbols: predictions from single star photoionisation models with {\em CoStar}
ionising fluxes for different ionisation parameters (Stasi\'nska \& Schaerer 1997).
Models with their ``standard''
$U$ are connected by the solid (dwarf stars) and dotted line (giants).
The predictions are shown for solar and 1/5 solar metallicity (left and right panel).}
\end{figure}

Due to the higher complexity (e.g.\ stratified ionisation, unknown hydrodynamics)
of WR atmospheres their modeling is presently more uncertain than that of O stars.
Although WR stars are less numerous than O stars, which usually dominate the 
ionising flux production, the former can in certain starburst phases 
contribute a non-negligible fraction of the ionising flux especially at high
energies (e.g.\ Schmutz \etal\ 1992, Schaerer \& Vacca 1998) and need thus to be 
investigated in a similar fashion.
It is presently not fully clear how realistic the hardness predicted by the
synthesis models is in these phases (e.g.\ Bresolin \etal\ 1999).
Regarding this issue it must be recognised that this prediction depends not only
on the atmosphere model used, but is also affected by the uncertainty in
the link between the interior and atmosphere in these phases (cf.\ Schmutz \etal\ 1992),
which is not discussed here.

The current status can be summarised as follows.
First, evidence for hard spectra of hot WR stars (WNE, WC/WO) at low metallicity 
comes from the presence of nebular \heii\ emission in some WR nebulae and in
many giant \hii\ regions harbouring WR stars (Garnett \etal\ 1991, Schaerer \etal\ 1999b)
and the rather good quantitative agreement between predicted and observed
nebular \Heii\ (Schaerer 1996, de Mello \etal\ 1998).
Indirectly this is also supported by the 
observations of O~{\sc iv} in IR spectra of two metal-poor galaxies 
(Schaerer \& Stasi\'nska 1999).
Second, there is evidence against hard WR spectra at higher metallicities
($\sim$ LMC -- solar): few nebulae with \heii\ emission are known;
line blanketed WR model atmospheres with softer spectra are in better agreement
with the properties of WR nebulae (Esteban \etal\ 1993, Crowther \etal\ 1999);
the observed EL ratios of \hii\ regions may be in conflict with photoionisation
models incorporating the Schmutz \etal\ (1992) WR atmospheres and
tracks from Maeder (1990; see Bresolin \etal\ 1999).
The latter finding is, however, strongly model dependent and no such discrepancy
is found in the models of Stasi\'nska \etal\ (2000) using up-to-date stellar tracks.
Somewhat surprisingly also, unblanketed WR atmosphere models are quite successfully
applied to PN with WR central stars (Pe\~na \etal\ 1998).
 
It is thus well possible that the pure-He models of Schmutz \etal\ (1992)
provide a reasonable description of the ionising fluxes of WR stars at low
metallicity, while at LMC, solar and higher metallicities the spectra 
should be softer due to line blanketing.
New developments of line blanketed WR atmospheres are in progress
(see IAU Symp.\ 193, van der Hucht \etal\ 1999). This work should be
of particular importance for the interpretation of young ``metal-rich''
starbursts.

\subsection{Starbursts in the optical and IR}
We shall now summarise recent work using combined evolutionary synthesis and
photoionisation models to study the properties of starbursts from their optical and
IR emission lines.

On one hand extensive model grids have been compared to samples of \hii\ galaxies
(e.g.\ Garc\ii a-Vargas \etal\ 1995, Stasi\'nska \& Leitherer 1996, Stasi\'nska 
\etal\ 2000). In particular such studies have shown that at subsolar metallicities
no variations of the IMF with O/H are necessary, contrary to earlier claims.
Instead of studying global trends of EL ratios, the modeling of individual objects, 
for which sufficient observational constraints are available, is a complementary 
approach providing in particular more insight to fundamental physical processes
responsible for the emission lines.
We shall now briefly discuss few such recent studies.

The giant \hii\ region NGC 7714 was modeled by Grac\ii a-Vargas \etal\ (1997) 
based on optical spectra (including also stellar features) and on morphological 
information. A composite population consisting of a young burst and an older
population including RSG was found. The main difficulties in reproducing 
the EL ratios concern \oia/\hbeta\ and \oiii\ $\lambda$4363/5007, which
are underpredicted by the model.
NGC 2363 was recently modeled by Luridiana \etal\ (1999) based on optical spectra
and H$\alpha$ imaging. They find that the strategic line ratios are not reproduced
by their model, if the metallicity derived from standard techniques is adopted.
They suggest that temperature fluctuations may be present which would lead
to an underestimate of O/H. Assuming a higher O/H leads to an agreement with
most observed line ratios. 
Interestingly, as for NGC 7714, the temperature sensitive ratio 
\oiii\ $\lambda$4363/5007 is underpredicted in all of their models\footnote{
Note: this is NOT a contradiction with their hypothesis of temperature 
fluctuations.}. 
Most clearly the problem of \oiiic\ is illustrated in the study of I Zw 18
by Stasi\'nska \& Schaerer (1999) using ground-based and HST UV--optical data.
The failure to reproduce the electron temperature deduced from \oiiic\ indicates
a missing energy source not included in the stellar photoionisation model. 
Although most observables can be reproduced by the combined starburst 
and photoionisation models --- and the tool can thus be used to derive
SB properties from the EL --- one has to conclude that for accurate
studies relying on nebular lines from \hii\ regions (and presumably also
more complex objects) some additional physical process(es) (possibly
shocks, conductive heating at X-ray interfaces etc.) must be taken 
into account. 

Analysis of IR observations (mostly from SWS and LWS on ISO) of starbursts 
based on combined SB + photoionisation models are just beginning to appear 
in the literature. In this context it is useful to keep some
intrinsic difficulties in mind are.
Given the nature of objects and the large apertures involved, the 
integrated spectrum generally includes a large variety of regions.
This fact, together with the complex geometries involved, render {\em a 
priori} the construction of photoionisation models difficult.

Simple models were constructed for case studies of Arp 299 and M82
by Satyapal \etal\ (1998) and Colbert \etal\ (1999) to interpret 
their LWS (40-200 \micron) spectra. Colbert \etal\ (1999) find
that the observed EL spectrum of M82 is compatible with an instantaneous
burst at ages $\sim$ 3--5 Myr, a Salpeter IMF, and a high upper mass
cut-off.
Surprisingly, inspection of models with similar ingredients (cf.\
Stasi\'nska \& Leitherer 1996), show that the shorter wavelength data 
(see Genzel \etal\ 1998) is clearly incompatible with
the Colbert \etal\ model predicting too hard a spectrum. In view
of the few line ratios originating from the \hii\ gas and the large
number of free parameters the photoionisation model is underconstrained.
A larger wavelength coverage or other constraints are required.

F\"orster-Schreiber (1998) has described the geometry of clusters and
gas clouds in M82 by a single ``effective'' ionisation parameter.
This value has been adopted as typical for a sample of 27 starbursts in 
the SB + photoionisation models of Thornley \etal\ (2000). Instead of
modeling a simple stellar population their models are based on an ensemble 
of \hii\ regions following an observed luminosity function, which overall
leads to a reduction, albeit small, of the hardness of the ionising 
spectrum. From the ISO/SWS  [Ne~{\sc iii}]/[Ne~{\sc ii}] line
ratios they conclude that the observations are compatible with a high
upper mass cut-off ($M_{\rm up} \sim$ 50--100 \msun). To reproduce 
the relatively low average  [Ne~{\sc iii}]/[Ne~{\sc ii}] ratio, short
timescales of SF are required.

A different approach has been taken by Schaerer \& Stasi\'nska (1999), who
modeled two well studied objects (NGC 5253, II Zw 40) with a fairly well
known massive star population and existing UV-optical-IR observations.
While their model successfully reproduces the stellar features and the 
observed ionisation structure of H, He, and O (as revealed from the optical
and IR lines), the predicted IR fine structure line ratios of
[Ne~{\sc iii}]/[Ne~{\sc ii}],  [Ar~{\sc iii}]/[Ar~{\sc ii}], and 
[S~{\sc iv}]/[S~{\sc iii}] show too high an excitation.
The origin of this discrepancy (atomic data? other?) is still unknown.
In any case this attempt to describe two relatively ``simple'' objects
illustrates the current limitations and shows that further progress
is needed for a proper understanding and use of the IR fine structure
lines as reliable diagnostics.
Improvement is expected from multi-wavelength analysis of simpler objects 
(e.g.\ Galactic and LMC \hii\ regions, PN) and other work. 
Such studies should be crucial to reliably extend the diagnostic tools
to the IR to fully exploit the enormous observational capabilities
provided by recent and upcoming facilities in probing the properties
of massive star formation from the local Universe to high reshift.

\begin{acknowledgements}
I thank the organisers for the invitation to this interesting
workshop and the LOC for financial support.
\end{acknowledgements}




\begin{references}
\reference Bresolin, F. , Kennicutt, R. C. , Jr., Garnett, D. R., 1999, \apj\ 
    510, 104 
\reference Crowther, P.A., et al., 1999, A\&A, 350, 1007
\reference Colbert, J.W., et al., 1999, \apj, 511, 521
\reference de Mello, D.F., Leitherer, C., Heckman, T.M., 2000, \apj, 530, 251
\reference de Mello, D.F., Schaerer, D., Heldmann, J., Leitherer, C., 1998, \apj, 507, 199
\reference Diaz, A.I., P\'erez-Montero, E., 2000, MNRAS 312, 130
\reference Esteban, C., 1993, A\&A, 272, 299
\reference F\"orster-Schreiber, N.M., 1998, PhD thesis, Ludwig-Maximilian-Universit\"at, Munich
\reference Garc\ii a-Vargas, M.L., \etal, 1995, A\&AS, 115, 13
\reference Garc\ii a-Vargas, M.L., \etal, 1997, ApJ, 478, 112
\reference Garnett, D.R., et al., 1991, \apj, 373, 458
\reference Genzel, R., \etal, 1998, \apj, 498, 579
\reference Gonz\'alez Delgado, R.M., P\'erez, E., 2000, \mnras, in press (astro-ph/0003067)
\reference Gonz\'alez Delgado, R.M., \etal, 1998, ApJ, 495, 698
\reference Gonz\'alez Delgado, R.M., \etal, 1999a, \apjs, 125, 479
\reference Gonz\'alez Delgado, R.M., \etal, 1999b, \apjs, 125, 489
\reference Guseva, N., Izotov, Y.I., Thuan, T.X., 2000, \apj, 531, 776
\reference Heap, S.R., 2000, in ``The evolution of Galaxies. I. Observational clues'', ApSS, 
  in press
\reference Heckman, T.M., et al., 1998, \apj, 503, 646
\reference Kennicutt, R.C., et al., 
    2000, ApJ, in press (astro-ph/0002180)
\reference Leitherer, C., 1999a, in ``Chemical evolution from zero to high redshift'', ESO workshop, 
   J.R. Walsh, M.R. Rosa (eds.), Berlin, Springer, p. 204
\reference Leitherer, C., 1999b, in ``Spectrophotometric Dating of Stars and Galaxies'', 
  I. Hubeny, S.R. Heap, R.H. Cornett (eds.), ASP Conf. Series 192, p. 3
\reference Leitherer, C., et al., 1999, ApJS, 123, 3
\reference Luridiana, V., \etal, 1999, ApJ, 527, 110
\reference Lowenthal, J.D., et al., 1997, \apj, 481, 673
\reference Maeder, A., 1990, \aaps, 84, 139
\reference Maeder, A., Meynet, G., 2000, ARAA, in press
\reference Mas-Hesse, J.M., 1999, in ``The Interplay between massive stars and the ISM'',
  JENAM99 meeting, D. Schaerer, R.M. Gonzalez Delgado (eds.), New Astronomy Reviews, in press
\reference Mas-Hesse, J.M., Kunth, D., 1999, \aap, 349, 765
\reference Oey, M.S., Kennicutt, R.C., 1997, MNRAS, 291, 827
\reference Oey, M.S., et al., 2000, \apjs, in press (astro-ph/9912363)
\reference Pe\~na, M., et al., 1998, A\&A, 337, 866
\reference Pauldrach, A., et al., 1998, in ``Boulder-Munich II'' Properties of hot, luminous stars'',
  I.D. Howarth (ed.), ASP Conf. Series, 131, p. 258
\reference Pettini, M., et al., 2000, \apj, 528, 96
\reference Satyapal, S., et al., 1998, in preparation
\reference Schaerer, D., 1996, \apj, 467, L17
\reference Schaerer, D., 1998,  in ``Boulder-Munich II'' Properties of hot, luminous stars'',
  I.D. Howarth (ed.), ASP Conf. Series, 131, p. 310
\reference Schaerer, D., 1999a,  in ``Spectrophotometric Dating of Stars and Galaxies'', 
  I. Hubeny, S.R. Heap, R.H. Cornett (eds.), ASP Conf. Series 192, p. 49
\reference Schaerer, D., 1999b, in ``Massive Stellar Clusters'', A. Lancon, C.M. Boily (eds.), 
   ASP Conf. Series, in press (astro-ph/0001531)
\reference Schaerer, D., Contini, T., Pindao M., 1999b, \aaps, 136, 35
\reference Schaerer, D., Contini, T., Kunth D., 1999a, \aap, 341, 399
\reference Schaerer, D., de Koter, A., 1997, \aap, 322, 598
\reference Schaerer, D., Guseva, N., Izotov, Y.I., Thuan, T.X., 2000, \aap,
   in press
\reference Schaerer, D., Stasi\'nska, G., 1999, \aap, 345, L17
\reference Schaerer, D., Vacca, W. D. 1998, \apj, 497, 618 
\reference Schmutz, W., Leitherer, C., Gruenwald, R., 1992, \pasp, 104, 1164
\reference Stasi\'nska, G., Leitherer, C., 1996, \apjs\ 107, 661
\reference Stasi\'nska, G., Schaerer, D., 1997, \aap, 322, 615
\reference Stasi\'nska, G., Schaerer, D., 1999, \aap, 351, 72
\reference Stasi\'nska, G., Schaerer, D., Leitherer, C., 2000, A\&A, in preparation
\reference Thornley, M.D., et al., 2000, \apj, in press (astro-ph/0003334)
\reference Vacca, W.D., Garmany, C.D., Shull, M.J., 1996, \apj, 460, 914
\reference van der Hucht, K., et al. (eds.), 1999, IAU Symp. 193
\reference Vilchez, J.M., Esteban, C., 1996, MNRAS 280, 720
\reference Vilchez, J.M., \etal, 1988, MNRAS, 235, 633
\end{references}
\end{document}